\documentclass[12pt,b5paper]{article}
\usepackage{subfigure}
\usepackage{amsmath,epsfig}

\usepackage{graphicx}% Include figure files
\usepackage{dcolumn}% Align table columns on decimal point
\usepackage{bm}% bold math

\begin{document}

\title{$K^{*}$ production in Heavy Ion Collisions at RHIC}

\author {Sadhana Dash(for the STAR Collaboration)\\ 
{\it Institute of Physics ,Bhubaneswar ,INDIA.}\\
e-mail: sadhana@iopb.res.in}
\date{}

\maketitle

\begin{abstract}
Study of resonances with their short life-times provides useful tools to
probe the properties of hot and dense matter produced in relativistic heavy ion
collisions. The high density and/or high temperature of the medium can modify
resonance properties such as mass and width. Therefore, measurement of these
properties can reveal important information about the evolution dynamics in
heavy ion collisions.
We report the measurements of $K^{*}$ transverse momentum($p_T$) spectra
at mid-rapidity via its hadronic decay channel up to intermediate $p_{T}$
of 2.9 GeV/c using the STAR detector in Au+Au and Cu+Cu collisions at
$\sqrt{s_{\mathrm{NN}}}$= 62.4 GeV and 200 GeV. These results are compared
to previously reported $K^{*}$ results from Au+Au collisions at RHIC.
Integrated yield ratios of $K^{*}/K$ and $K^{*}/\phi$ are used to understand 
the rescattering and regeneration effects on $K^{*}$ production.
\vspace{0.4in}
       PACS   PACS  : 12.38.Mh, 25.75.-q, 25.75.Dw  
\end{abstract}

\section{Introduction}
Ultrarelativistic heavy ion collisions are used to study the quantum
chromodynamics in the extreme conditions of high temperature and high energy
density\cite{whitepaper}. The $K^{*}$s have a very short life time 
($\sim 4fm/c$) which is comparable to the expected life time of the fireball 
created in heavy ion collisions\cite{haibinPRC,xin dong}. In the dense medium
created, resonances are produced in close proximity with other strongly 
interacting hadrons, and hence the in-medium effects related to the high
density/or high temperature of the medium can modify the characteristic 
properties such as masses, widths, and spectra shapes. Therefore, studies of 
resonance production mechanism can be a useful tool in understanding the 
properties of the high density matter\cite{haibinPRC}\cite{medium2}.

Due to the short life time, some of the $K^*$s produced at hadronization 
can decay in the medium before the kinetic freezeout, and are less likely to
be reconstructed due to the rescattering of the daughter particles.
Alternatively the interaction between the large population of $\pi$ and $K$,
may regenerate $K^*$ increasing the observed yield\cite{regeneration}. 
These two competing processes determine the final $K^{*}$ yield which depends
on the time elapsed between the chemical and kinetic freeze-out, the source
size and the interaction cross section of daughter hadrons.
Since the $\pi \pi$ total interaction cross section\cite{pipicross} is
significantly larger (about a factor $\sim 5$) than the $\pi K$ total 
interaction cross-section\cite{pik}, the final observable $K^*$ yield in
heavy ion collisions at RHIC is expected to be smaller than the primordial
yield. This should be evident in a suppression of the $K^*/K$ and/or
$K^*/\phi$ yield ratios in AA compared to elementary pp collisions at similar
collision energy. Comparison of those ratios can then be used to roughly 
estimate the lower limit on the time difference between chemical freeze-out
and kinetic freeze-out\cite{haibinPRC}. Centrality dependence of this
suppression can be used to gauge the size of the fireball. The system size 
and energy dependence studies can shed additional light on the different
in-medium effects, particularly the interplay between regeneration and 
rescattering mechanisms.

\section{Experiment and Data Analysis}

The primary tracking device, TPC(Time Projection Chamber) within STAR was used 
to measure the $K^*$ production via its hadronic decay channel. TPC provides
the particle identification and the momentum information of the charged 
particles by measuring their ionisation energy loss(dE/dx)\cite{tpc}.
The results discussed here are taken from Au+Au at $\sqrt{s_{\mathrm{NN}}}$=
62.4 GeV and Cu+Cu collisions at $\sqrt{s_{\mathrm{NN}}}$= 200 GeV and
$\sqrt{s_{\mathrm{NN}}}$= 62.4 GeV at RHIC.

Charged kaons and pions were selected from the primary tracks whose distance
of closest approach to the primary vertex have values less than 1.5 cm.
The kaons and pions were selected requiring the dE/dx to be within two
standard deviations(2$\sigma$) from the Bethe-Bloch expectation for energy 
loss. Both the kaons and pions were required to have atleast 15 fit points
for the tracks reconstructed inside the TPC and the ratio of the number of fit
points to the number of maximum possible fit points was greater than 0.55 to 
avoid selecting split tracks. Further kaon and pion tracks were selected with 
both momentum and $p_{T}$ greater than 0.2 GeV/c.
  
The unlike-sign K$\pi$ invariant mass distribution was reconstructed from
random combination of pairs from an event. The combinatorial background 
distribution was built by using mixed-event technique where the unlike-sign 
K$\pi$ invariant mass was obtained from different events\cite{eventmix}.
The data sample was divided into 10 bins in multiplicity and $V_{z}$ position.
The pairs from events in same multiplicity and vertex position bins were
selected for mixing to ensure that the event characteristic remain similar 
between different events. The mixed event generated was normalized to subtract 
the background in the same event unlike-sign invariant mass spectrum. The 
normalisation factor was evaluated by taking the ratio between the number of
entries in the unlike-sign same event and mixed event distributions for
invariant mass greater than 1.1 GeV/c as K$\pi$ pairs with invariant mass
greater than 1.1 GeV are less likely to be correlated.  
After subtracting the normalized mixed event background from the unlike
sign spectrum, we observe the $K^{*}$ signal. In the unlike sign spectrum we
also have higher and/or lower K$\pi$ mass resonant states and nonresonant
correlations due to particle misidentification which contribute significantly
to the residual correlations near the signal. These attributes are not present 
in the mixed event spectrum and thus the mixed event only removes the
uncorrelated background pairs from the unlike sign spectrum.

\begin{figure*}[htp]
\begin{minipage}{0.4\textwidth}
\centering
\includegraphics[height=13pc,width=14pc]{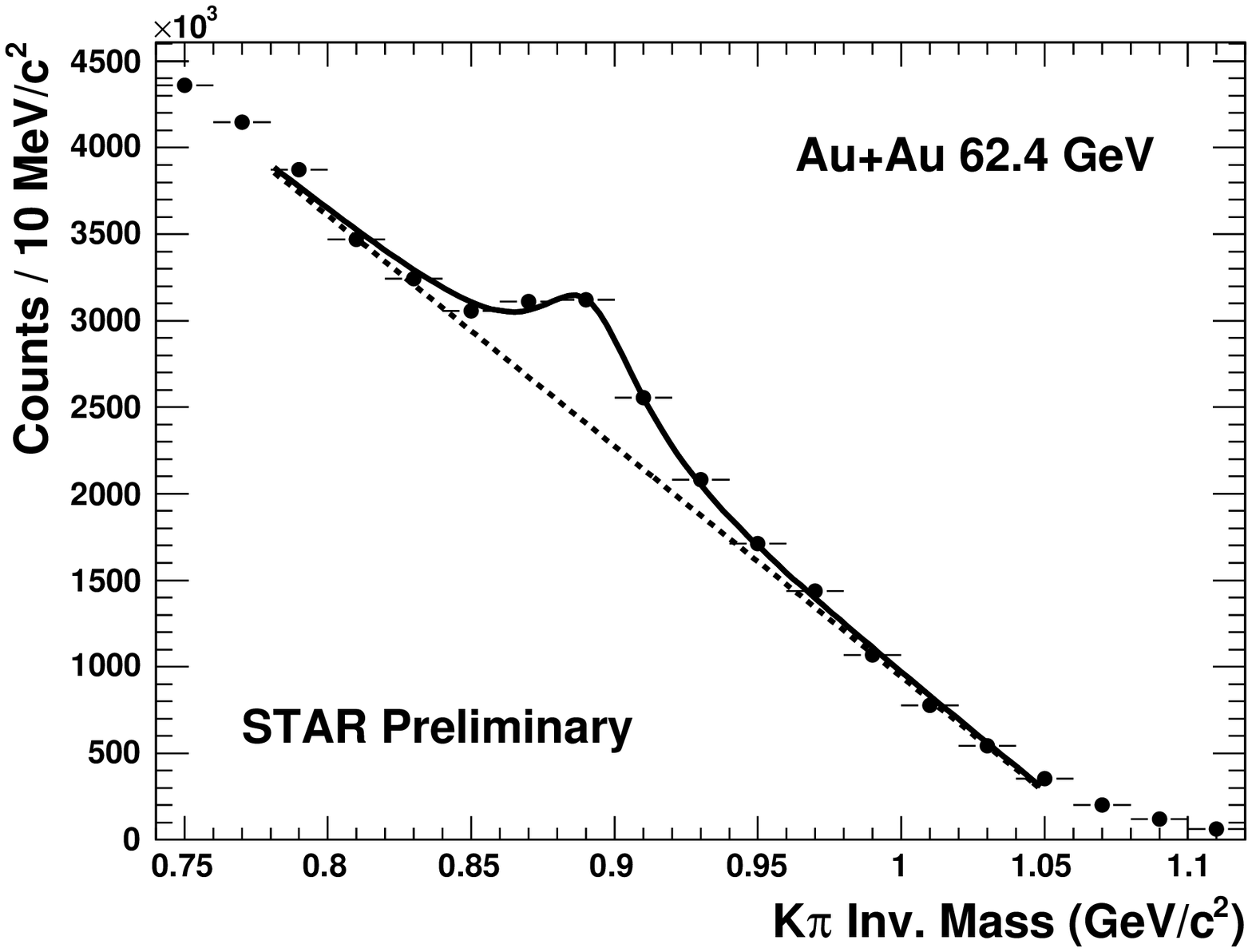}
\end{minipage}
\begin{minipage}{0.4\textwidth}
\centering
\includegraphics[height=13pc,width=14pc]{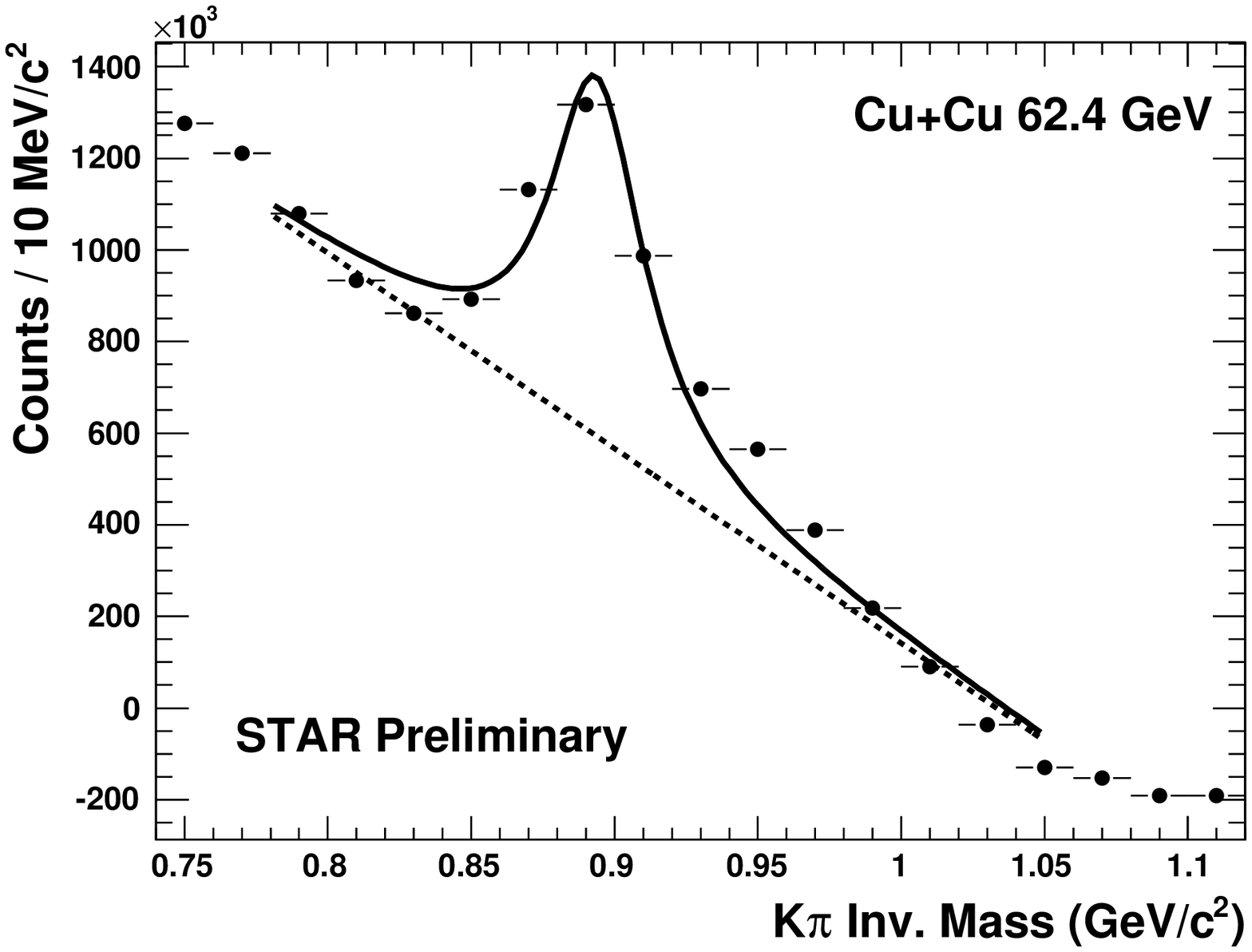}
\end{minipage}
\caption{The K$\pi$ pair invariant mass spectrum after mixed-event
background subtraction fitted to SBW + RB. Left panel: Au+Au at 62.4 GeV
Right panel: Cu+Cu at 62.4 GeV}
\label{signal}
\end{figure*}

\section{Results}

\begin{figure*}[htp]
\begin{minipage}{0.45\textwidth}
\centering
\includegraphics[height=13pc,width=14pc]{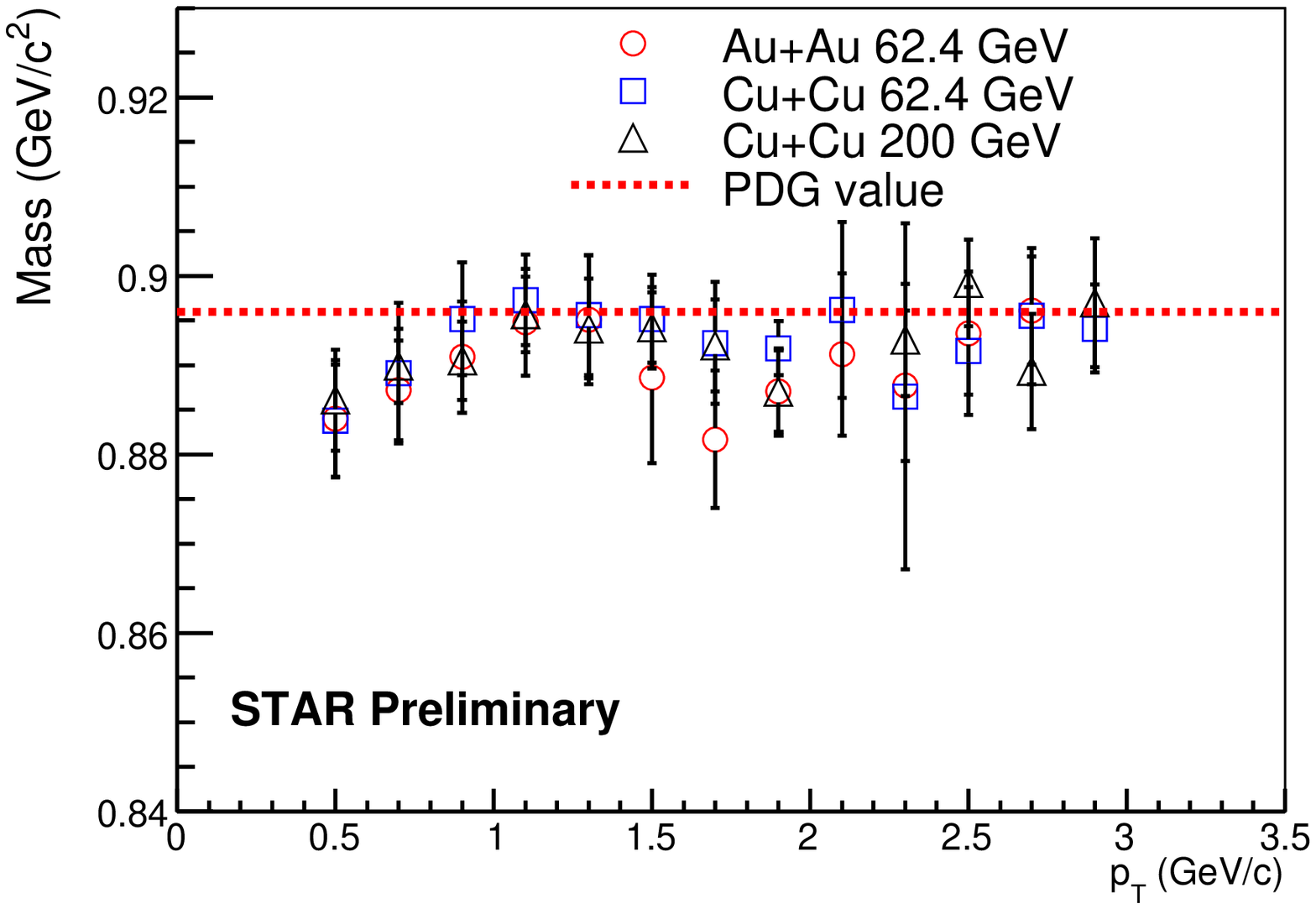}
\end{minipage}
\begin{minipage}{0.45\textwidth}
\centering
\includegraphics[height=13pc,width=14pc]{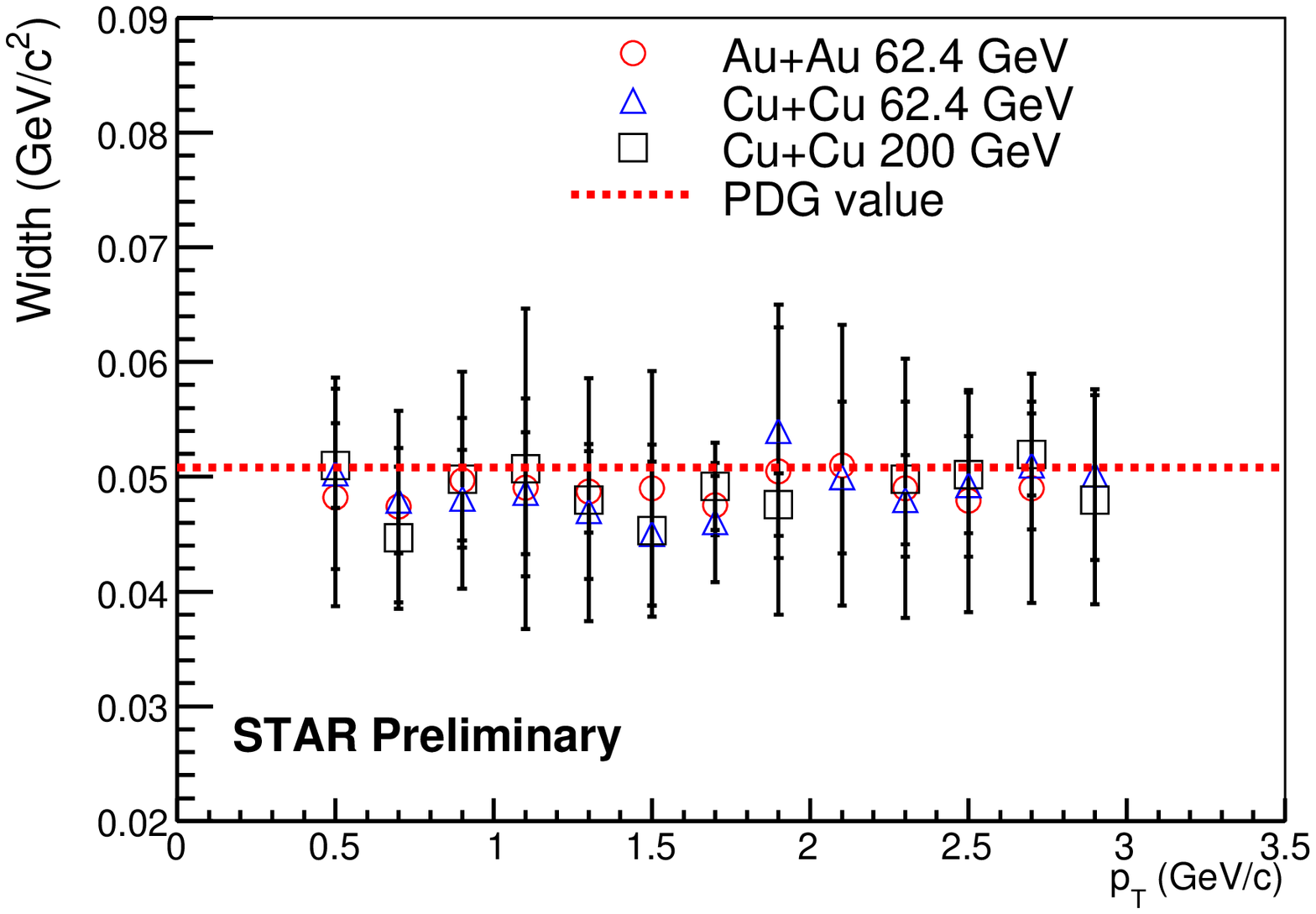}
\end{minipage}
\caption{$K^{*0}$ mass (left panel) and width (right panel) as a function of
 $p_{T}$ in minimum bias Au+Au collisions at 62.4 GeV and Cu+Cu collisions 
at 200 and 62.4 GeV. The solid line stand for the PDG values.}
\label{masswidth}
\end{figure*}

Figure \ref{signal} shows the unlike sign K$\pi$ invariant mass spectrum
after normalized mixed event background subtraction in Au+Au and Cu+Cu
collisions at $\sqrt{s_{\mathrm{NN}}}$= 62.4 GeV. 
The invariant mass distribution is fitted to the function SBW + RBW where
SBW is the non relativistic Breit wigner function and RBG is the linear
function describing the residual background\cite{haibinPRC}.
The variation of $K^{*0}$ mass and width with respect to $p_{T}$ for minimum
bias Au+Au and Cu+Cu collisions is shown in Figure \ref{masswidth}.
Within the errors, we do not observe a large difference in the measured $K^{*0}$ mass from the PDG\cite{pdg} value for most of the $p_{T}$ range studied. The 
$K^{*0}$ width measured is comparable to the PDG value.The systematic
uncertainities in the mass and width were calculated by varying the particle
types, background subtraction functions and the dynamical cuts.

\begin{figure*}[htp]
\begin{minipage}{0.4\textwidth}
\centering
\includegraphics[height=13pc,width=14pc]{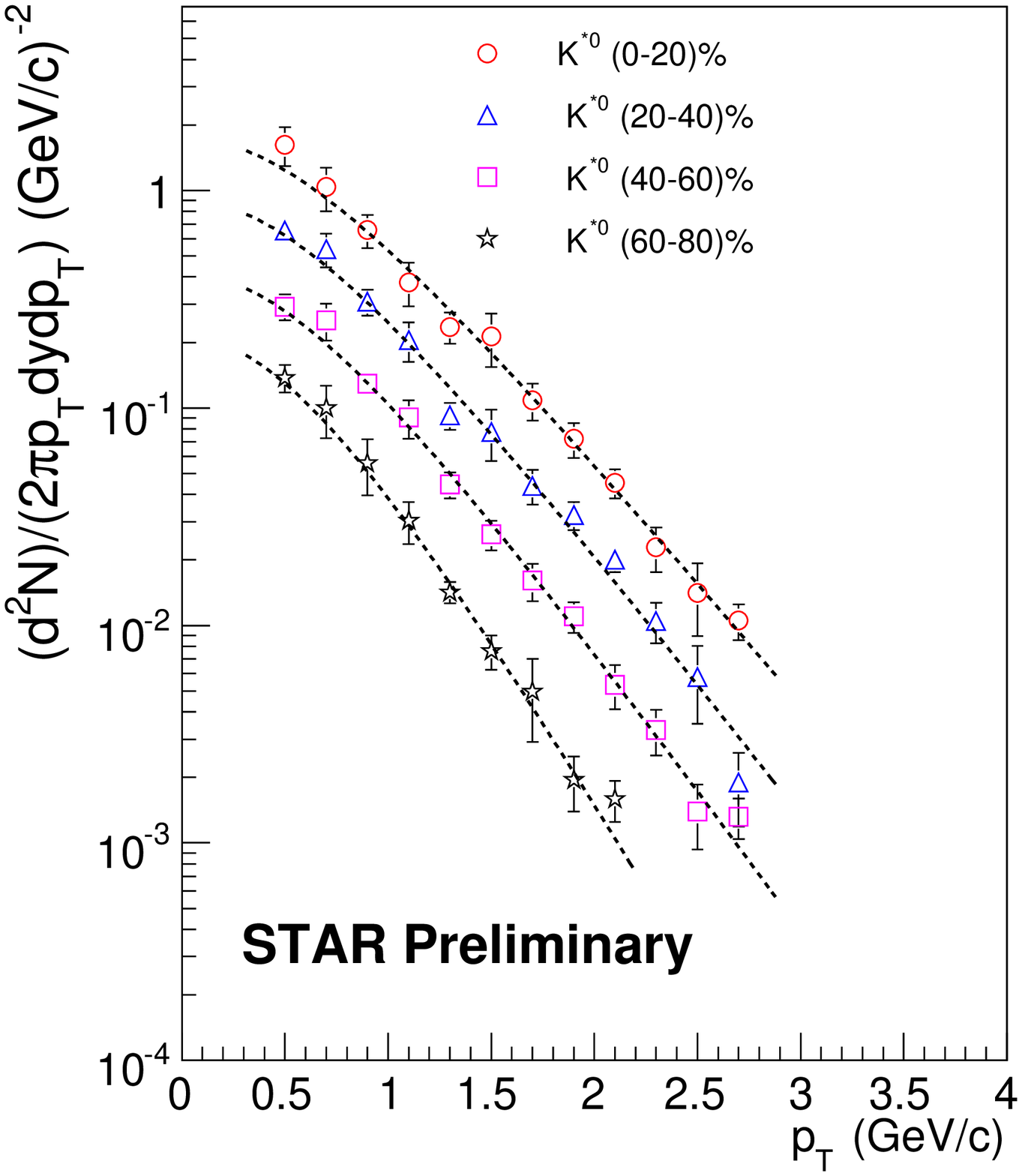}
\end{minipage}
\begin{minipage}{0.4\textwidth}
\centering
\includegraphics[height=13pc,width=14pc]{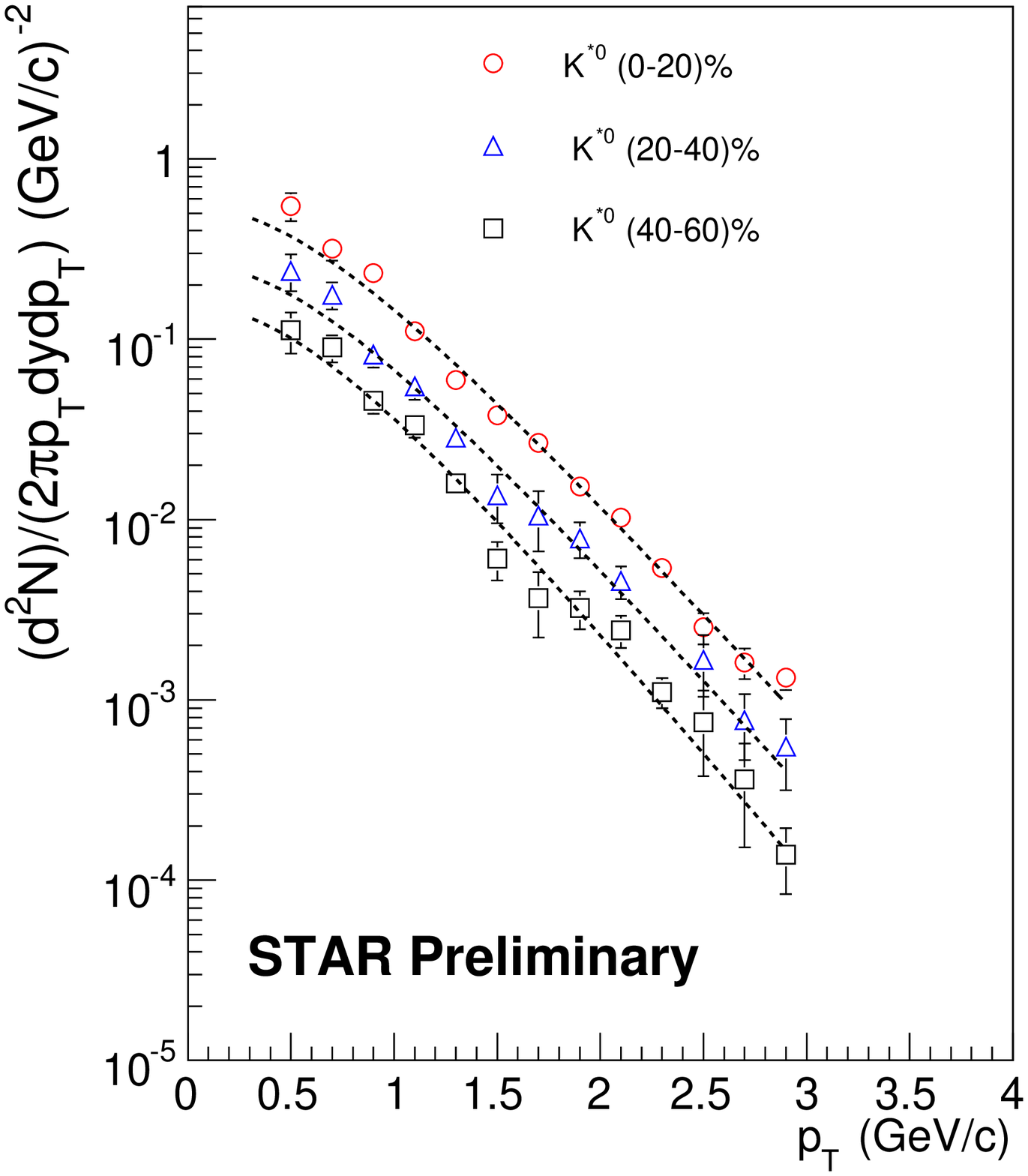}
\end{minipage}
\caption{$p_{T}$ spectra in Au+Au(left panel) and Cu+Cu(right panel)collisions
at 62.4 GeV. The dashed line represents the exponential fit to data.}
\label{spectra}
\end{figure*}

Figure \ref{spectra} shows the $K^{*}$ mid-rapidity transverse momentum spectra
for different centralities of Au+Au and Cu+Cu collisions at $\sqrt{s_{\mathrm{NN}}}$= 62.4 GeV. The raw spectra yields,  corrected for efficiency and 
detector acceptance are well described by exponential fit function.

\begin{figure*}[htp]
\begin{minipage}{0.4\textwidth}
\centering
\includegraphics[height=13pc,width=14pc]{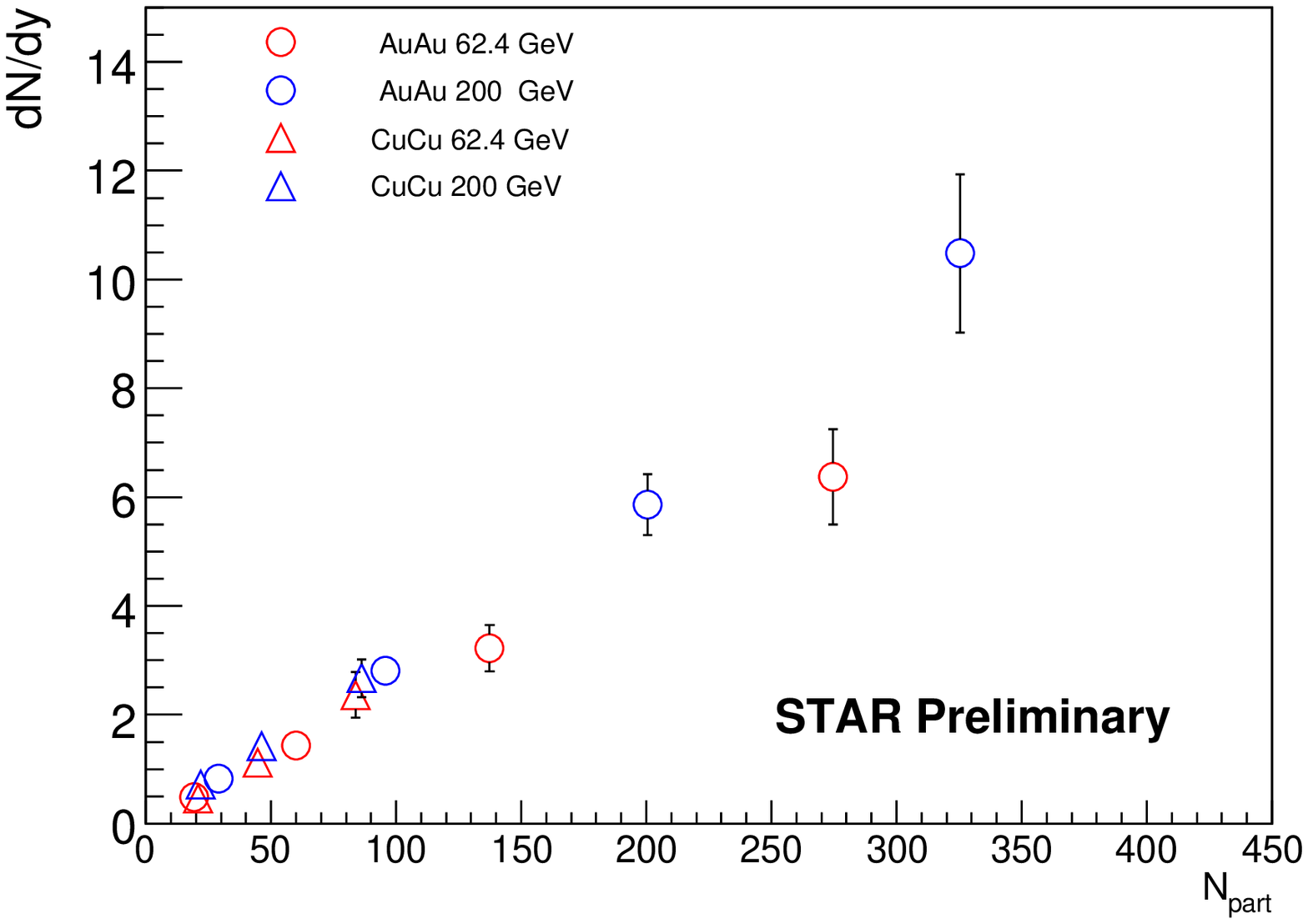}
\end{minipage}
\begin{minipage}{0.4\textwidth}
\centering
\includegraphics[height=13pc,width=14pc]{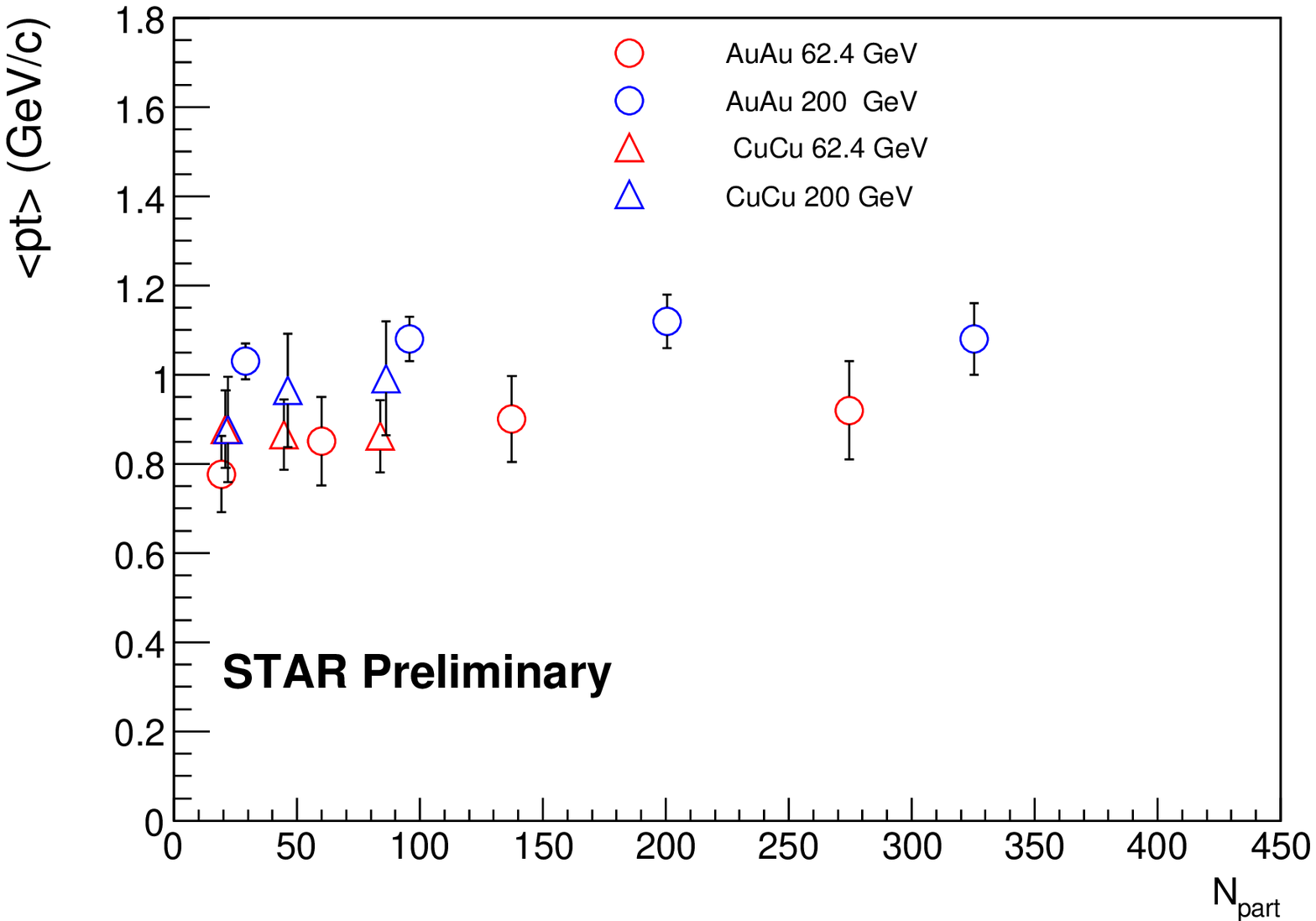}
\end{minipage}
\caption{$K^{*0}$ invariant yield (left panel) and mean $p_{T}$(right panel)as
a function of number of participants}
\label{meanptdndy}
\end{figure*}

The $K^{*0}$ yield at midrapidity is calculated from the data points in the
measured range and exponential fit was used to extract the yield outside the
fiducial range. The $K^{*0}$ invariant yield increases with number of
participants in both Au+Au and Cu+Cu collisions. The systematic uncertainities
on $K^{*0}$ dN/dy is estimated by using different fit functions, particle types
and dynamical cuts. The variation of measured $K^{*}$ dN/dy with respect to
number of participants is depicted in Figure \ref{meanptdndy}.
The $K^{*0}$ mean $p_{T}$ was evaluated using the data points in the measured
range of the $p_{T}$ spectrum while assuming an exponential behaviour outside
the fiducial range. The systematic uncertainity includes the differences 
between the direct calculation and from all other sources mentioned earlier.
The $K^{*0}$ mean $p_{T}$ as a function of number of participants is shown in
Figure \ref{meanptdndy} (right panel) for all the collision systems discussed.
No significant centrality and system size dependence of mean $p_{T}$ is
observed for $K^{*0}$ in Au+Au and Cu+Cu collisions at a given colliding 
energy.

\begin{figure*}[htp]
\begin{minipage}{0.45\textwidth}
\centering
\includegraphics[height=13pc,width=14pc]{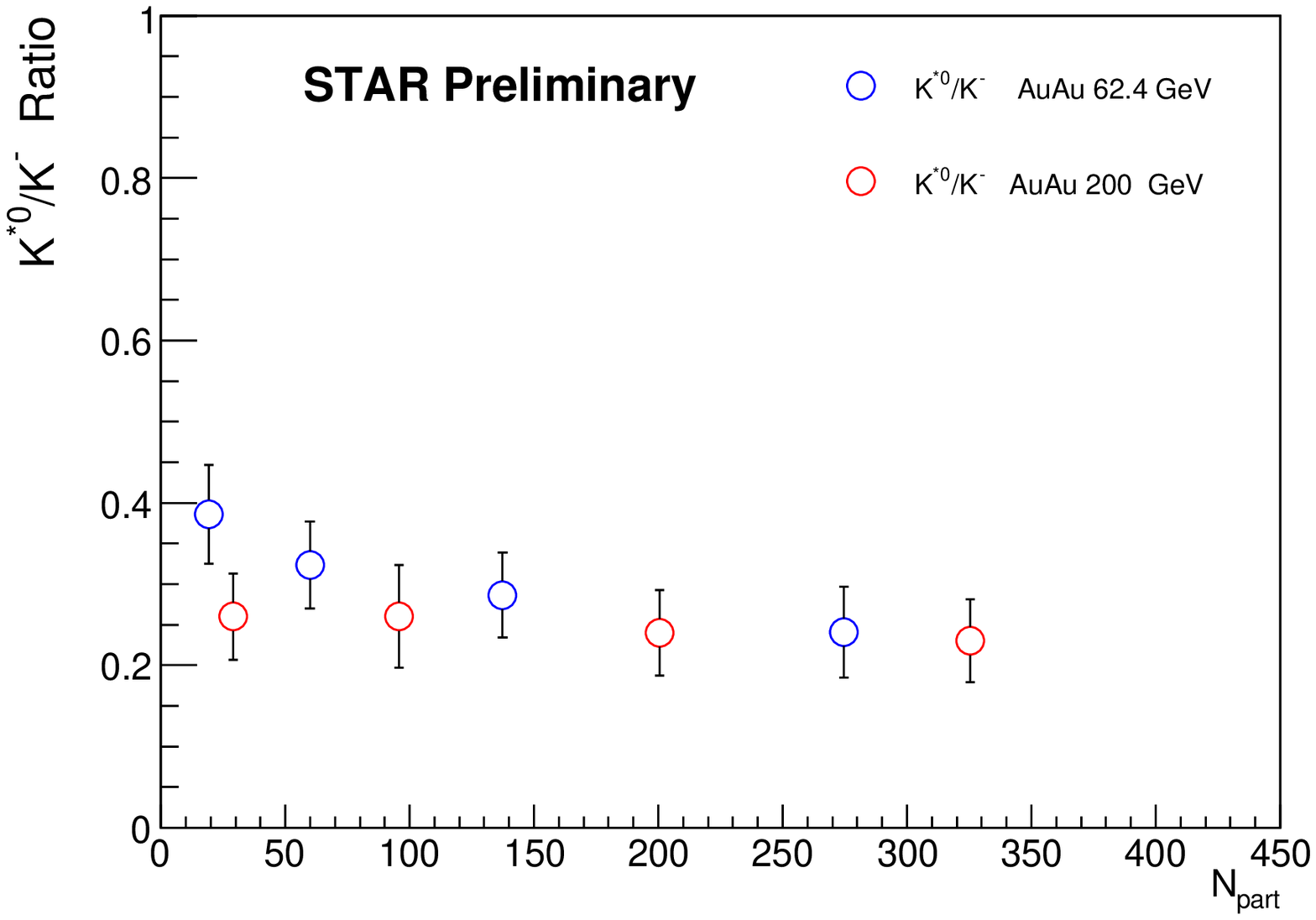}
\end{minipage}
\begin{minipage}{0.45\textwidth}
\centering
\includegraphics[height=13pc,width=14pc]{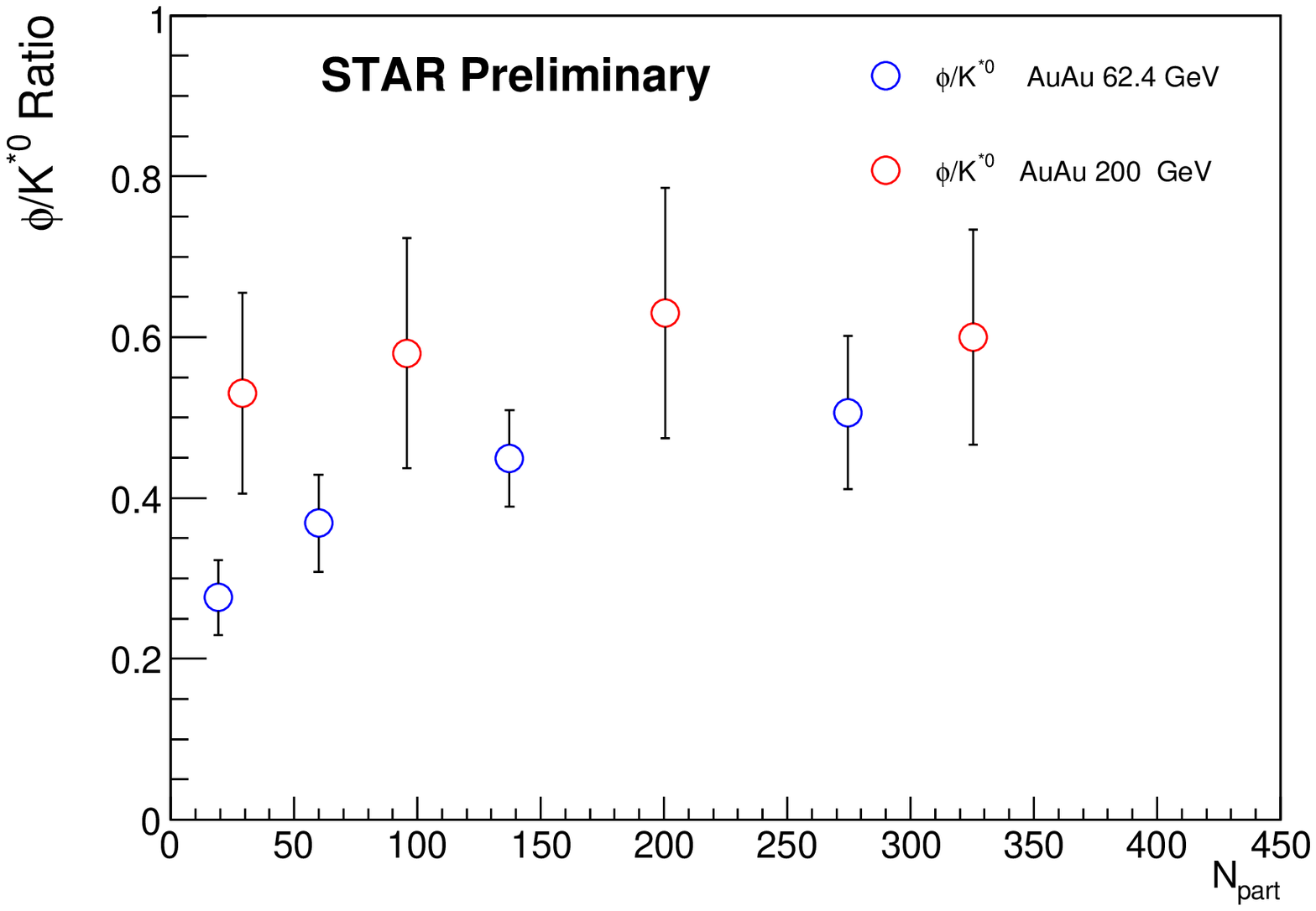}
\end{minipage}
\caption{$K^{*0}/K^{-}$ ratio (left panel)and $\phi/K^{*0}$ ratio(right panel)
as a function of number of participants.}
\label{particleratio}
\end{figure*}

The measurement of $K^{*0}/K^{-}$ yield ratio can provide vital information 
on the $K^{*}$ production properties as $K^{*0}$ and $K^{-}$ have different 
masses and spin but identical quark content. This ratio was observed to be 
smaller in Au+Au collisions compared to p+p collisons at the same beam energy
\cite{haibinPRC,xin dong}. Figure \ref{particleratio}(left panel) shows that
 $K^{*0}/K^{-}$ ratio decreases with number of participants in Au+Au collisions
at 62.4 and 200 GeV. This may indicate that the rescattering effect is dominant
over the regeneration effect and the fireball created in central collisions has
comparatively larger lifetime than the peripheral collisions. 
Similarly $\phi/K^{*0}$ ratio may also give us an idea on rescattering and 
regeneration effect as both $\phi$ amd $K^{*0}$ have same spin and similar
mass but different strangeness number and lifetime.
In Figure \ref{particleratio}(right panel), we observe that $\phi/K^{*0}$ ratio
increases with number of participants in Au+Au collisons at 62.4 and 200 GeV 
favouring the dominance of rescattering effect. Since $\phi$ and $K^{*0}$ have
different strangeness number, the observed increase may be due to the possible 
strangeness enhancement in more central collisons.

\section{Summary}
The preliminary results on the $K^{*}$ production in Au+Au and Cu+Cu collisions
measured with the STAR detector at RHIC at $\sqrt{s_{\mathrm{NN}}}$ = 62.4 GeV
and  $\sqrt{s_{\mathrm{NN}}}$ = 200 GeV are presented. The $K^{*}$ integrated
yield increases with number of participants in both Au+Au and Cu+Cu collision
systems. No significant dependence of mean $p_{T}$ on system size is observed.
The particle ratio measurement highlights the dominance of rescattering effect
over the regeneration mechanism in $K^{*}$ production in heavy ion collisions.

\end{document}